\documentclass[12pt,preprint]{aastex}

\slugcomment{ApJ Letters accpeted}
\shorttitle{Early optical afterglow of GRB041006}
\shortauthors{Urata et al.} 
\begin{document}
\title{Very early multi-color observations of the plateau phase of GRB 041006 afterglow}

\author{
Y \textsc{Urata}\altaffilmark{1,2}, 
K.Y. \textsc{Huang}\altaffilmark{3}, 
Y.L. \textsc{Qiu}\altaffilmark{4},
J. \textsc{Hu}\altaffilmark{4},
P.H. \textsc{Kuo}\altaffilmark{3},
T. \textsc{Tamagawa}\altaffilmark{2}, 
W.H. \textsc{Ip}\altaffilmark{3},
D. \textsc{Kinoshita}\altaffilmark{3},
H. \textsc{Fukushi}\altaffilmark{5},
M. \textsc{Isogai}\altaffilmark{6},
T. \textsc{Miyata}\altaffilmark{6},
Y. \textsc{Nakada}\altaffilmark{6},
T. \textsc{Aoki}\altaffilmark{6}, 
T. \textsc{Soyano}\altaffilmark{6},
K. \textsc{Tarusawa}\altaffilmark{6},
H. \textsc{Mito}\altaffilmark{5},
K. \textsc{Onda}\altaffilmark{1}
M. \textsc{Ibrahimov}\altaffilmark{7}
A. \textsc{Pozanenko}\altaffilmark{8}
and
K. \textsc{Makishima}\altaffilmark{2,9} 
}

\altaffiltext{1}{Department of Physics, Saitama University, Shimo-Okubo, Saitama,  338-8570, Japan}
\altaffiltext{2}{RIKEN (Institute of Physical and Chemical Research), 2-1 Hirosawa, Wako, Saitama 351-0198, Japan}
\email{urata@heal.phy.saitama-u.ac.jp}
\altaffiltext{3}{Institute of Astronomy, National Central University, Chung-Li 32054, Taiwan, Republic of China}
\altaffiltext{4}{National  Astronomical Observatories, Chinese Academy of Sciences, Beijing 100012, China}
\altaffiltext{5}{Institute of Astronomy, University of Tokyo, 2-21-1 Osawa, Mitaka, Tokyo 181-0015, Japan}
\altaffiltext{6}{Kiso Observatory, Institute of Astronomy, The University of Tokyo, Mitake-mura, Kiso-gun, Nagano 397-0101, Japan}
\altaffiltext{7}{Ulugh Beg Astronomical Institute, Tashkent 700052, Uzbekistan}
\altaffiltext{8}{Space Research Institute, 84/32 Profsoyuznaya Str, Moscow 117997, Russia}
\altaffiltext{9}{Department of Physics, University of Tokyo, 7-3-1 Hongo, Bunkyo-ku, Tokyo 113-0033, Japan}

\begin{abstract}
   
Observations of the optical afterglow of GRB 041006 with the Kiso
Observatory 1.05 m Schmidt telescope, the Lulin Observatory 1.0 m
telescope and the Xinglong Observatory 0.6 m telescope. Three-bands
($B$, $V$ and $R$) of photometric data points were obtained on 2004
October 6, 0.025$-$0.329 days after the burst.  These very early multi
band light curves imply the existence of a color dependent plateau
phase.  The $B$-band light curve shows a clear plateau at around 0.03
days after the burst. The $R$ band light curve shows the hint of a
plateau, or a possible slope change, at around 0.1 days after the
burst.  The overall behavior of these multi-band light curves may be
interpreted in terms of the sum of two separate components, one
showing a monotonic decay the other exhibiting a rising and a falling
phase, as described by the standard afterglow model.

\end{abstract}


\section{Introduction}

Based on the standard afterglow model (e.g. Sari, Piran and Narayan
1998), an optical light curve can be expected to consist of a
combination of four power-law segments, with connections at certain
break frequencies.  The break frequencies are the self-absorption
frequency ($\nu_{sa}$), the typical frequency ($\nu_m$) and the
cooling frequency ($\nu_{c}$).  Before the {\it HETE-2} era, almost
all GRB light curves showed only the $\nu > \nu_c$ segment
characterized by a single declining or achromatic break. Note that the
standard afterglow model predicts the existence of a peak at an early
time when the typical synchrotron $\nu_{m}$ frequency crosses into the
optical frequency. Before the peak time $(\nu_{opt} < \nu_m)$, the
luminosity increases proportional to $t^{1/2}$, until reaching the
maximum flux $F_{\rm max}$ at $\nu=\nu_{m}$; the subsequent decay of
the luminosity is proportional to $t^{3(1-p)/4}$, where $p$ is the
index of the power-law distribution of energetic electrons accelerated
at the shock.  Since the break time varies with the frequency,
multi-color observations of the very early afterglow are required, so
as to catch $\nu_{m}$ in the optical wavelength. Up to now, there have
only been a few cases (such as GRB 021004; Urata et al 2005a) for
which early time multi-color data are available.

GRB 041006 was detected with {\it HETE-2} at 12:18:08 UT on 2004
October 06.  The Wide-Field X-ray Monitor (WXM; Shirasaki et al 2003)
localized the burst in real time, resulting in a GCN alert 42 seconds
after the burst trigger.  The flight error region was a circle with a
$14'$ radius (90\% confidence level) centered at
$\alpha^{2000}=00^{\rm h}54^{\rm m}54^{\rm s}$,
$\delta^{2000}=01^{\circ}18' 37''$.  Spectral analysis showed the 2-30
keV fluence of this event to be $5\times10^{-6} {\rm erg/cm^{2}}$ and
the 30-400 keV fluence to be $7\times10^{-6} {\rm erg/cm^{2}}$: the
ratio between these two values means that it can be classified as an
``X-ray rich GRB''. The light curve shape of the gamma-ray pulse of
GRB 041006 is very similar to that of GRB 030329 but, its spectral
characteristics are 20 times fainter.  GRB 041006 shows a soft
precursor before the main gamma-ray pulse (Galassi et al. 2004).
At 1.4 hours after the burst, the optical afterglow was found within
the $14'$-radius error circle at the following coordinates:
$\alpha^{2000}=00^{\rm h}54^{\rm m}50^{\rm s}.17$,
$\delta^{2000}=01^{\circ}14' 07''$ (Da Costa et al. 2004).  The
redshift was determined by Price et al. (2004) using the
Gemini North telescope to be $z=0.716$.

\section{Observations}

\begin{figure}[htb]
\begin{center}
\includegraphics[width=0.9\textwidth]{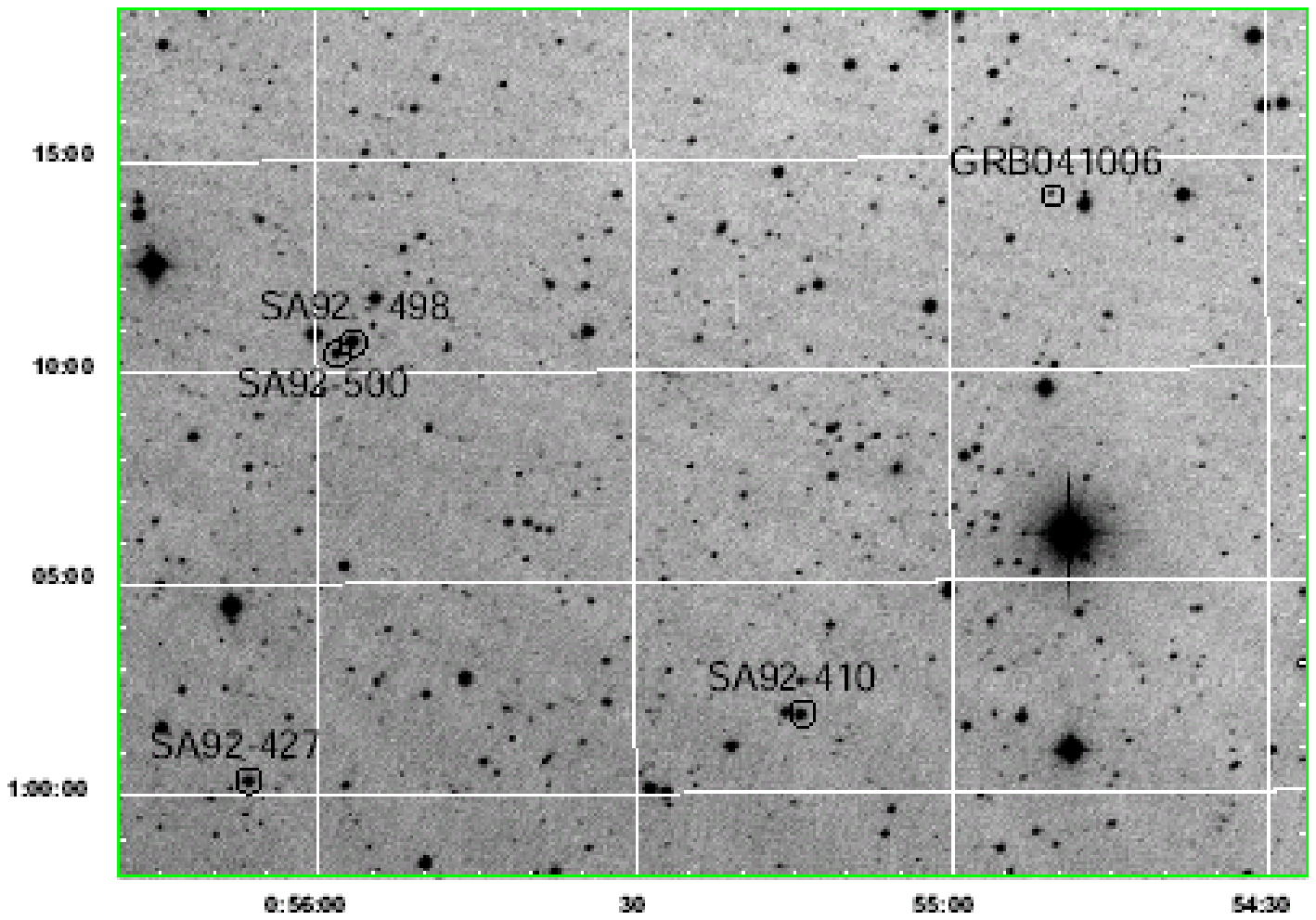}
\caption{An $R$ band image of the GRB 041006 field taken at the Kiso 
observatory. The circles indicate the afterglow and standard
stars. North is up; east is left.}
\label{grb041006image}
\end{center}
\end{figure}

The follow-up observations of the GRB 041006 optical afterglow at
the Kiso, Lulin and Xinglong Observatories were carried out within the
framework of the East Asia Follow-up Observation Network (EAFON, Urata
et al. 2005b). The 1.05 m Schmidt telescope and a $2k\times2k$ CCD
camera at the Kiso observatory were used to make 300 sec exposure $B$,
$V$, and $R$ band imaging observations, starting at 12:51 UT on 2004
October 6 (0.009 days after the burst). The field of view is
$51'.2\times51'.2$, and the pixel size is $1''.5$ square.  Before
receiving the report by Da Costa et al. (2004), we had attempted to
cover the entire {\it HETE-2} error region. We made $B$, $V$, and $R$
band imaging observations of 300 sec exposure each. The afterglow was
clearly detected in all bands. In order to image simultaneously the
optical afterglow and several Landolt standard stars in SA92 (Landolt
1992), we pointed the telescope 5 arcmin to the east (Figure
\ref{grb041006image}).

The $R$ band observations were performed using the 0.6 m telescope at
the Xinglong Observatory, starting at 2004 October 6 (14:33 UT; 0.094
days after the burst). The filed of view is $10'.1\times 10'.1$, and
the pixel size is $0''.47$ square.  The $B$ and $R$ band observation
of the afterglow was also performed with the Lulin-One-meter-Telescope
\citep{lot} on the night of October 6 ($0.320-0.329$ days after the
burst).  Additional $B$, $V$, $R$ and $I$ band observations were made
by the Maidnak 1.5 m telescope on October 6.  The time coverage of the
observations was from 0.149 to 0.188 days after the burst.

\section{Analysis}

The data reduction was carried out using the standard package NOAO
IRAF package. We performed the bias-subtraction and flat-fielding
correction using the appropriate calibration data. We calibrated the
flux zero-point of our images through a comparison with the SA92
standard star field. We also performed cross calibrations of our
photometric results using several field stars for which the magnitudes
have been calibrated by Henden (2004). The difference of photometric
zero points between our SA92 and Henden's filed photometry is within
0.04 mag.  Aperture photometry for all the data was performed using
the APPHOT package of IRAF.  The photometric results are summarized in
Table \ref{grb041006obs}.

\section{Results}

\subsection{Light curves}

\begin{figure}[htb]
\begin{center}
\includegraphics[width=0.9\textwidth]{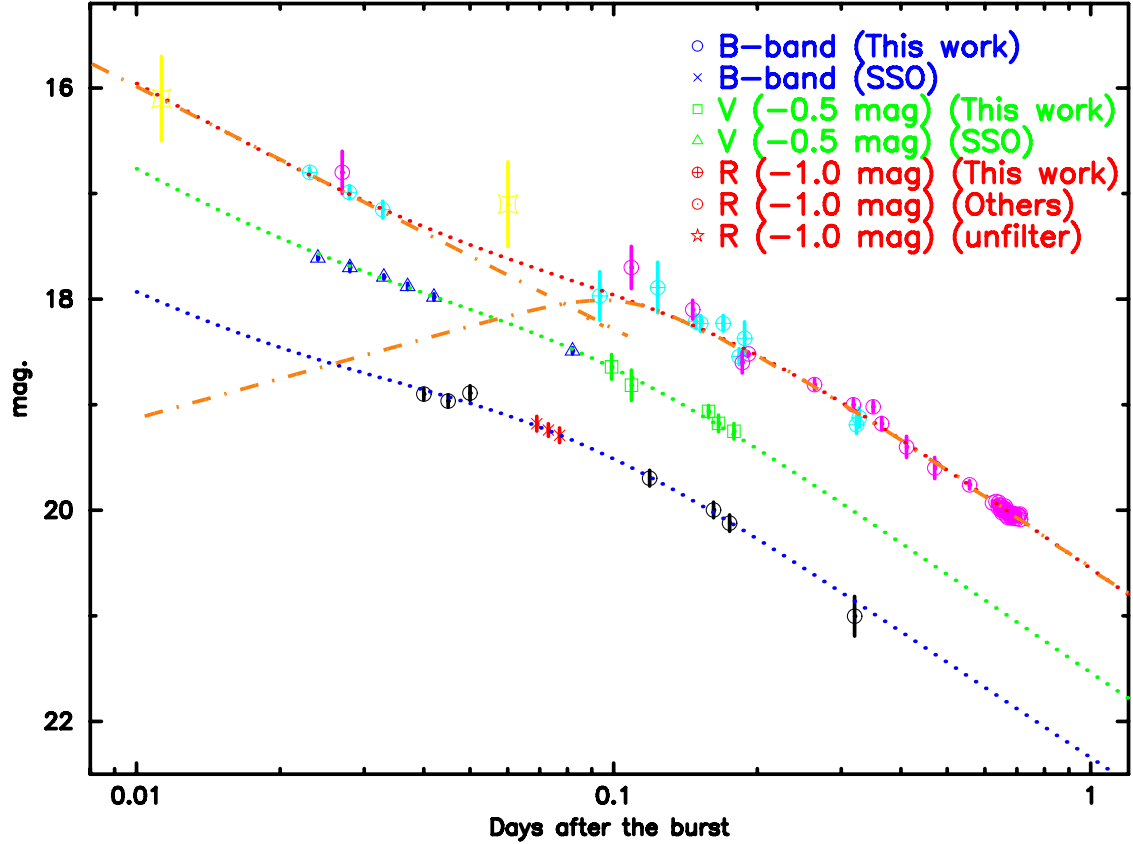}
\caption{$B$, $V$ and $R$ band light curves produced based on the Kiso, Lulin and Beijing results together with SSO\cite{soderberg},
MMT\cite{stanek} and several GCNs.  The dotted lines indicate the best
fit model functions described in the text. The dashed lines indicate
the model components of the best fit function for the $R$ band
lightcurve.}
\label{grb041006lc}
\end{center}
\end{figure}

The multi-band light curves of the GRB 041006 afterglow are shown in
Figure \ref{grb041006lc}. In addition to our $B$, $V$ and $R$ band
data, there are several unfiltered observations, the $R$ band data
points reported in the GCN Circulars (Yost et al. 2004; Ayani \&
Yamaoka 2004; Fugazza et al. 2004; Monfardini et al. 2004; Misra \&
Pandey 2004; Kahharov et al. 2004; Kinugasa \& Torii 2004), and a
number of published $R$ and $V$ band data points (Soderberg et
al. 2005; Stanek et al. 2005) were included in Figure 2. The $B$-band
light curve showed a clear plateau at around 0.03 days after the
burst, although its behavior in earlier epochs remains
unknown. Interestingly, the $R$ band light curve shows a hint of a
plateau, or a possible slope change, around 0.1 days after the burst.
These observational results indicate that the $B$, $V$ and $R$ band
photometric points obtained by EAFON play an important role in
characterizing the temporal evolution of the afterglow.

As is obvious in the Figure \ref{grb041006lc}, the single power law
fit ($\alpha=-0.95$, $\chi^{2}/\nu=5.76$ with $\nu$=56) of the $R$
band light curve is poorly described. It is also unlikely that the
$V$ band light curve can be described with the single power law
($\alpha = -0.72$, $\chi^{2}/\nu=1.54$ with $\nu$=9).  On the other
hand, the data subsets at $t<0.04$ days and $t>0.15$ days can be
described successfully by two separate single power-law models.  For
the $R$ band, we have obtained $\alpha= -0.91\pm0.12$ with
$\chi^{2}/\nu= 0.30$ for $\nu= 2$ before 0.04 days, and $\alpha=
-1.12\pm0.02$ with $\chi^{2}/\nu =1.44$ for $\nu =48$ after 0.15 days;
for the $V$ band $\alpha= -0.59\pm0.03$ with $\chi^{2}/\nu= 0.41$ for
$\nu= 3$ before 0.04 day, and $\alpha= -1.41\pm0.30$ with
$\chi^{2}/\nu =0.19$ for $\nu =1$ after 0.15 days From this point of
view, the overall behavior of these multi band light curves may then
be understood as the sum of two separate components, one showing a
monotonic decay with the other having a rising and a falling phase.

\subsection{Color change of the afterglow}

The photometric results were corrected for Galactic reddening, using
the reddening map of Schlegel, Finkbeiner, \& Davis (1998).  The
Galactic reddening toward the burst is $E(B-V$)= 0.022, which implies
a Galactic extinction of $A_{B}= 0.094$ and $A_{R}= 0.058$.  Although
the light curves in Figure \ref{grb041006lc} suggest a significant
color change within $<0.1$ days, our data coverage is insufficient to
make a conclusion about such a possibility. Accordingly, for
simplicity's sake we assume that at $t>0.15$ days the power-law
spectrum of the afterglow has a constant index.  The observed color,
$B-R=0.83\pm0.28$ mag, then indicates a spectral index of
$\beta\sim-1.24\pm0.42$ in terms of $f(\nu)\propto \nu^{\beta}$. This
is close to the value of $\beta=-1.07\pm0.02$, which is predicted by
the spherically symmetric model of Sari, Piran, \& Halpern (1999) in
the regime of $\nu_{c}<\nu_{opt}$ in combination with the measured
$\alpha=-1.118\pm0.016$.

\section{Discussion}

We observed the GRB 041006 optical afterglow from the very early phase
($\sim0.03$ days after the burst) in three bands.  Although the
relatively early phase of a dozen afterglows such as GRB 990123
\cite{akerlof}, multi-color observations of very early optical
afterglow are still rare.  In the GRB 021004 case, the very early
optical afterglow shows a clear achromatic re-brightening phase which
peaks was at around $\sim0.07$ days \citep{kiso}. On the other hand,
the afterglow behavior of GRB 041006 depends on its color. The $B$
band light curve shows a clear plateau at around 0.03 days after the
burst, although its behavior in earlier epochs remains unknown. The
$R$ band light curve shows the hint of a plateau, or a possible slope
change, around 0.1 days after the burst.

Several models, based on the standard afterglow model, can be used to
explain the variability in the light curve, such as the variable
external density (e.g. Lazati et al. 2002), refreshed shock (Rees \&
M\'esz\'aros 1998; Kumar \& Piran 2000a; Sari \& M\'esz\'aros 2000),
and patchy shell models (Kumar \& Piran 2000b). To explain the present
multi-band light curves, these models require similar temporal
behavior in each band. If the light curve of GRB 041006 does not show
a color change, it should have two bumps with almost the same
amplitudes. The patchy shell model is not suitable to explain current
light curves in this case. This is because the model predicts that
the amplitude of the fluctuations in the afterglow light cure is
expected to decrease with time proportional to the Lorentz factor.

The suggested second component may be identified with the brightening
phase of Sari et al's standard afterglow model (1998) which predicts
the evolution to consist of four phases involving a convex-shaped
period; the brightening phase can be expressed with a power-law index
of 0.5.  To describe the plateau phase together with the
monotonically decreasing first component, we use the single power law,
plus a smoothly broken power law function expressed as
\begin{eqnarray}
F_{\nu}(t) =\frac{F^{*}_{\nu}}{[(t/t_{b})^{\alpha_{1}}+(t/t_{b})^{\alpha_{2}}]},
\end{eqnarray}
with $\alpha_{1}$ fixed at 0.5 after Sari et al. (1998).  We obtain an
acceptable fit for the $R$ band light curve with $\alpha=-1.07\pm0.01$,
$\alpha_{2}=-1.49\pm0.03$ and $t_{b}=0.143\pm0.012$
($\chi^{2}/\nu=1.19$ with $\nu=66$).  We also fit the $B$ and $V$ band
light curves with the same function. For better constraints, we fix
the $\alpha_{1}=-1.0$ for $B$ band and $\alpha_{2}=-1.4$ of the $V$
band. Due to the lack of data points for the first components in the
$B$ band, and the former components in the $V$ band., these vales come
from the result of $R$ band fitting.  For the $B$ band, we obtained
$\alpha_{2}=-1.47\pm0.07$ and $t_{b}=0.0675\pm0.005$
($\chi^{2}/\nu=0.47$ with $\nu=7$); for the $V$ band,
$\alpha=-1.08\pm0.02$ and $t_{b}=0.073\pm0.006$ ($\chi^{2}/\nu=0.83$
with $\nu=8$).  In Figure \ref{grb041006lc}, these best fit functions
are superposed on the $B$, $V$ and $R$ band data points.  Based on the
equation (Sari et al. 1998)

\begin{equation}
\nu_{m}  =  5.7 \times 10^{14} \epsilon_B^{1/2} \epsilon_e^2
                             E_{52}^{1/2} t_d^{-3/2} {\rm \ Hz},
\end{equation}
we can estimate the strength parameter of the magnetic field $\epsilon_{B}$
and/or the injected electrons parameter $\epsilon_{e}$ as follows;
$\epsilon_{B}^{1/2}\epsilon_{e}^{2}=4.1\times10^{-3}$. This value is
consistent with the result of Panaitescu \& Kumar (2001, 2002) which
was estimated from late-time multi-frequency afterglow observations.

About the first component, the decay index is $\alpha=-0.92$ and the
back-extrapolated power-law function can successfully account for the
point at 14.4 min after the burst observed by ROTSE-III
\citep{yost}. The temporal index is considered to be directly related
to the prompt emission, as observed in the case of GRB 990123
\citep{akerlof} or GRB 030418 \citep{rykoff}. However, the temporal
index is inconsistent with the values ($-0.5$ or $-2$) predicted by a
reverse shock model \citep{kobayashi}. Thus the first component is not
likely to be an optical flash, such as is the case for GRB 990123.
Considering the above discussions one of possible explanation is that
these emissions come from multiple jets and/or sub-jets (Nakamura
2000, Yamazaki et al. 2004). In the multiple jet case, the former
emissions and the latter components are caused by narrower and wider
jets, respectively.

\section{Conclusion}

Thanks to our EAFON observations, we have obtained early B, V and R
band possible re-brightening episode for GRB 041006. While there are
several other afterglows similar to GRB 050319 (Wozniak et al 2005;
Huang et al 2006) and GRB 050525a \citep{klotz} for which the early
optical behaviors may possibly be similar to that of GRB 041006, their
physical mechanism is not clear, due to the general lack of
color/spectrum information on the early optical afterglows.  The
present work has provided some hints for the current shallow decaying
of the X-ray and optical behaviors.

\acknowledgments 

We thank the staff and observers at the Kiso, Lulin and XingLong
observatories for the various arrangements.  Y.U acknowledges support from
the Japan Society for the Promotion of Science (JSPS) through JSPS
Research Fellowships for Young Scientists. This work is also partly
supported by grants NSC 94-2752-M-008-001-PAE, NSC 94-2112-M-008-002,
and NSC 94-2112-M-008-019.


\begin{table}[htb]
\begin{center}
\caption{Log of follow-up observations of the afterglow of GRB 041006.\label{grb041006obs}}
\begin{tabular}{ccccccc}
\tableline\tableline
Date & Start Time(UT) & Delay (days) &Filter & Exposure (s) & mag & Site\\
\tableline
2004-10-06 & 13:13:04 & 0.040 & $B$ & 300 s $\times$ 1 & 18.900$\pm$0.055 & Kiso  \\
2004-10-06 & 13:20:05 & 0.045 & $B$ & 300 s $\times$ 1 & 18.965$\pm$0.058 & Kiso  \\
2004-10-06 & 13:27:05 & 0.050 & $B$ & 300 s $\times$ 1 & 18.890$\pm$0.070 & Kiso  \\
2004-10-06 & 15:07:22 & 0.119 & $B$ & 300 s $\times$ 1 & 19.696$\pm$0.074 & Kiso  \\
2004-10-06 & 16:09:02 & 0.162 & $B$ & 300 s $\times$ 1 & 19.997$\pm$0.074 & Mt. Maidanak  \\
2004-10-06 & 16:27:48 & 0.175 & $B$ & 300 s $\times$ 1 & 20.122$\pm$0.078 & Mt. Maidanak  \\
2004-10-06 & 19:57:08 & 0.320 & $B$ & 300 s $\times$ 1 & 21.004$\pm$0.188 & Lulin \\ 
\hline
2004-10-06 & 14:38:52 & 0.099 & $V$ & 300 s $\times$ 1 & 19.141$\pm$0.117 & Kiso  \\
2004-10-06 & 14:52:56 & 0.109 & $V$ & 300 s $\times$ 1 & 19.316$\pm$0.145 & Kiso  \\ 
2004-10-06 & 16:03:40 & 0.158 & $V$ & 240 s $\times$ 1 & 19.564$\pm$0.057 & Mt. Maidanak  \\
2004-10-06 & 16:14:57 & 0.166 & $V$ & 240 s $\times$ 1 & 19.677$\pm$0.078 & Mt. Maidanak  \\
2004-10-06 & 16:33:28 & 0.179 & $V$ & 300 s $\times$ 1 & 19.752$\pm$0.074 & Mt. Maidanak  \\ \hline
2004-10-06 & 12:51:20 & 0.025 & $R$ & 300 s $\times$ 1 & 17.799$\pm$0.040 & Kiso  \\
2004-10-06 & 12:58:21 & 0.030 & $R$ & 300 s $\times$ 1 & 17.988$\pm$0.063 & Kiso  \\
2004-10-06 & 13:05:23 & 0.035 & $R$ & 300 s $\times$ 1 & 18.152$\pm$0.080 & Kiso  \\
2004-10-06 & 14:32:51 & 0.094 & $R$ & 120 s $\times$ 1 & 18.97$\pm$0.23   & Beijing \\
2004-10-06 & 15:16:15 & 0.124 & $R$ & 120 s $\times$ 1 & 18.89$\pm$0.24   & Beijing \\
2004-10-06 & 15:49:32 & 0.149 & $R$ & 180 s $\times$ 1 & 19.211$\pm$0.067 & Mt. Maidanak  \\
2004-10-06 & 15:54:56 & 0.152 & $R$ & 180 s $\times$ 1 & 19.231$\pm$0.075 & Mt. Maidanak  \\
2004-10-06 & 16:19:59 & 0.170 & $R$ & 180 s $\times$ 1 & 19.229$\pm$0.074 & Mt. Maidanak  \\
2004-10-06 & 16:39:30 & 0.183 & $R$ & 300 s $\times$ 1 & 19.544$\pm$0.076 & Mt. Maidanak  \\
2004-10-06 & 16:46:28 & 0.189 & $R$ & 300 s $\times$ 1 & 19.371$\pm$0.156 & Mt. Maidanak  \\
2004-10-06 & 20:02:54 & 0.324 & $R$ & 300 s $\times$ 1 & 20.349$\pm$0.090 & Lulin \\
2004-10-06 & 20:09:25 & 0.329 & $R$ & 300 s $\times$ 1 & 20.314$\pm$0.092 & Lulin \\ \hline
\tableline	      	      	  	         	  
\end{tabular}		      
\end{center}		      
\end{table}		      

\end{document}